# New Iron Arsenide Oxides $(Fe_2As_2)(Sr_4(Sc,Ti)_3O_8)$, $(Fe_2As_2)(Ba_4Sc_3O_{7.5})$, and $(Fe_2As_2)(Ba_3Sc_2O_5)$

Naoto Kawaguchi[1,2], Hiraku Ogino[1,2*], Yasuaki Shimizu[1,2], Kohji Kishio[1,2], and Jun-ichi Shimoyama[1,2]

[1]Department of Applied Chemistry, The University of Tokyo, 7-3-1 Hongo, Bunkyo, Tokyo 113-8656, Japan

[2]Japan Science and Technology Agency-Transformative Research-project on Iron Pnictides (JST-TRIP), Sanban, Chiyoda, Tokyo 102-0075, Japan

We synthesized new layered iron arsenide oxides $(Fe_2As_2)(Sr_4(Sc,Ti)_3O_8)$, $(Fe_2As_2)(Ba_4Sc_3O_{7.5})$, and $(Fe_2As_2)(Ba_3Sc_2O_5)$. The crystal structures of these compounds are tetragonal with a space group of *I*4/*mmm*. The structure of $(Fe_2As_2)(Sr_4(Sc,Ti)_3O_8)$ and $(Fe_2As_2)(Ba_4Sc_3O_{7.5})$ consists of the alternate stacking of antifluorite $Fe_2As_2$ layers and triple perovskite-type oxide layers. The interlayer distance between the Fe planes of $(Fe_2As_2)(Ba_4Sc_3O_{7.5})$ is ~18.7 Å. Moreover, the *a*-axis of $(Fe_2As_2)(Ba_3Sc_2O_5)$ is the longest among the layered iron pnictides, indicating the structural flexibility of the layered iron pnictide containing perovskite-type layers. The bulk sample of $(Fe_2As_2)(Sr_4(Sc_{0.6}Ti_{0.4})_3O_8)$ exhibited diamagnetism up to 28 K in susceptibility measurements.

*Email address: tuogino@mail.ecc.u-tokyo.ac.jp

Since the discovery of high-$T_c$ superconductivity in LaFeAs(O,F)[1], a large number of layered compounds having an antifluorite-type Fe*Pn* (*Pn* = pnictide) layer have been discovered. Among these compounds, the series of pnictide oxides having perovskite-type oxide layers[2-7] are attractive because of their chemical flexibility, particularly of the perovskite-type oxide layer, which may result in a new superconductor family. These compounds can be classified into two structural groups according to the crystal structure of the perovskite-type oxide layers. $(Fe_2As_2)(Sr_3Sc_2O_5)$ (22325 phase)[2] is a representative compound having a double-perovskite layer between FeAs layers. On the other hand, $(Fe_2P_2)(Sr_4Sc_2O_6)$ ($T_c$ = 17 K)[3], $(Fe_2As_2)(Sr_4M_2O_6)$ ($M$ = Sc, Cr[4], V[5], MgTi[6]), and $(Fe_2As_2)(Ba_4Sc_2O_6)$[7] (22426 phase) have blocking layers regarded as the $K_2NiF_4$-type structure. The crystal structures of these phases are shown in Figs. 1(a) and 1(b). These new compounds indicate that the perovskite-type layer has a large variety of constituent elements. Thus, the perovskite-type layer has both chemical and structural flexibility. Besides these structures, chalcogenide oxides $(Cu_2Ch_2)(Sr_4Mn_3O_{7.5})$ (22438 phase) ($Ch$ = S, Se)[8] have a layered structure with a triple perovskite-type layer as shown in Fig. 1(c), where the oxygen content is assumed to be 8.

In the present study, we have discovered new iron pnictide oxides having the 22438 structure, $(Fe_2As_2)(Sr_4(Sc,Ti)_3O_8)$ and $(Fe_2As_2)(Ba_4Sc_3O_{7.5})$. In addition, a new 22325 compound $(Fe_2As_2)(Ba_3Sc_2O_5)$ was found.

All samples were synthesized by solid-state reaction starting from FeAs(3N), SrO(2N), $Sc_2O_3$(4N), Ti(3N), $TiO_2$(3N), Ba(2N), and $BaO_2$(3N). Starting compositions were fixed according to the general formula $(Fe_2As_2)(Sr_4(Sc_{1-x}Ti_x)_3O_8)$ ($0.4 \leq x \leq 0.6$), $(Fe_2As_2)(Ba_4Sc_3O_{7.5})$, $(Fe_2As_2)(Ba_3Sc_2O_5)$, or slightly FeAs-poor compositions. Since the starting reagents, SrO and Ba, are sensitive to moisture or oxygen in air, manipulations were carried out under an inert gas atmosphere in a glove box. Powder mixtures were pelletized and sealed in evacuated quartz ampoules, and reacted at 1000 ~1250°C for 40~120 h. Constituent phases were determined by powder X-ray diffraction (XRD) using a Rigaku Ultima-IV diffractometer and intensity data were collected in the 2θ range of 4°–80° at intervals of 0.02° using Cu-$K_\alpha$ radiation. Silicon powder was used for the internal standard. Magnetic susceptibility was evaluated by a superconducting quantum interference device (SQUID) magnetometer (Quantum Design MPMS-XL5s). Electric resistivity was measured by the AC four-point-probe method using a Quantum Design Physical Properties Measurement System.

The powder XRD patterns of $(Fe_2As_2)(Sr_4(Sc_{1-x}Ti_x)_3O_8)$ ($x$ = 0.4, 0.5, 0.6), $(Fe_2As_2)(Ba_4Sc_3O_{7.5})$, and $(Fe_2As_2)(Ba_3Sc_2O_5)$ reacted at 1200°C for 40 h are shown in Fig. 2. In $(Fe_2As_2)(Sr_4(Sc_{1-x}Ti_x)_3O_8)$, 22438 phase formed as the main phase at $x$ = 0.4~0.6, while weak peaks due to impurities, such as $SrTiO_3$ and FeAs, were always observed. Similar to the case for $(Fe_2As_2)(Sr_4MgTiO_6)$[6)], no superstructure corresponding to the ordering of Sc and Ti was observed from the powder XRD patterns of these samples. On the other hand, $(Fe_2As_2)(Ba_4Sc_3O_{7.5})$ and $(Fe_2As_2)(Ba_3Sc_2O_5)$ samples were composed of the 22438 or 22325 phase as the major phase, with small amount of FeAs. Note that $(Fe_2As_2)(Ba_3Sc_2O_5)$ was formed only when 20~30% FeAs-poor starting compositions were used. No peaks due to the $SrFe_2As_2$ or $BaFe_2As_2$ phases were found in any of the samples. The space group of all three compounds is $I4/mmm$. No structural change was observed for any compound down to 50 K. The lattice constants of $(Fe_2As_2)(Sr_4(Sc_{1-x}Ti_x)_3O_8)$ are $a$ = 4.024 Å, $c$ = 35.546 Å, $a$ = 4.010 Å, $c$ = 35.294 Å, and $a$ = 4.006 Å, $c$ = 35.032 Å for $x$ = 0.4, 0.5, and 0.6, respectively.

Systematic decreases in *a*- and *c*-axis lengths are reasonable because the ionic radius of $Ti^{4+}$ (0.605 Å) is less than that of $Sc^{3+}$ (0.730 Å). The *a*- and *c*-axis lengths of $(Fe_2As_2)(Ba_4Sc_3O_{7.5})$ are 4.123 and 37.565 Å, respectively, and those of $(Fe_2As_2)(Ba_3Sc_2O_5)$ are 4.133 and 28.355 Å, respectively. The *a*-axis of $(Fe_2As_2)(Ba_3Sc_2O_5)$ is the longest among the layered iron pnictides. In the iron arsenide oxides with a perovskite-type layer, the *a*-axis length can be controlled over a wide range by changing constituent elements in the blocking layer from 3.918 Å for $(Fe_2As_2)(Sr_4Cr_2O_6)$[4)] to 4.133 Å for $(Fe_2As_2)(Ba_3Sc_2O_5)$. This indicates that iron pnictides with a thick perovskite-type blocking layer have the advantages of not only chemical and structural flexibility but also structural control of the pnictide layer.

The mean valence of *M* in $(Fe_2As_2)(AE_4M_3O_8)$ is calculated to be 3.3. Therefore, partial substitution of tetravalent Ti for trivalent Sc is necessary to form $(Fe_2As_2)(Sr_4(Sc_{1-x}Ti_x)_3O_8)$. On the other hand, the 22438 phase forms without partial Ti substitution when *AE* = Ba. As reported for $(Cu_2Ch_2)(Sr_4Mn_3O_{7.5})$[8)], generation of an oxygen vacancy is considered to decrease the valence of *M* to +3, resulting in the generation of $(Fe_2As_2)(Ba_4Sc_3O_{7.5})$. Although 22438 phase was obtained from starting compositions of $(Fe_2As_2)(Ba_4Sc_3O_y)$ ($y$ = 7.5 and 8), phase purity was slightly better for a sample with $y$ = 7.5.

Temperature dependences of zero-field-cooled (ZFC) and field-cooled (FC) magnetization of $(Fe_2As_2)(Sr_4(Sc_{1-x}Ti_x)_3O_8)$ ($x$ = 0.4, 0.5, 0.6) and $(Fe_2As_2)(Ba_4Sc_3O_{7.5})$ are shown in Fig. 3. The bulk samples of $(Fe_2As_2)(Sr_4(Sc_{0.6}Ti_{0.4})_3O_8)$, $(Fe_2As_2)(Sr_4(Sc_{0.5}Ti_{0.5})_3O_8)$, and $(Fe_2As_2)(Ba_4Sc_3O_{7.5})$ had superconducting transitions at 28, 22, and 11 K, respectively. Although the superconducting volume fraction calculated from ZFC magnetization at 2 K under 1 Oe was ~13% for $(Fe_2As_2)(Sr_4(Sc_{0.6}Ti_{0.4})_3O_8)$, it increased to more than 20% under 0.2 Oe. This fact and the small difference between ZFC magnetization and FC magnetization indicates the granular superconducting nature of $(Fe_2As_2)(Sr_4(Sc_{0.6}Ti_{0.4})_3O_8)$. The volume fraction decreased with an increase in the Ti composition and was almost negligible for $x = 0.6$. From the magnetization of $(Fe_2As_2)(Ba_4Sc_3O_{7.5})$, it is difficult to conclude that the observed small diamagnetism is due to superconductivity of this phase.

Figures 4(a)–4(c) show the temperature dependence of the resistivities for $(Fe_2As_2)(Sr_4(Sc_{1-x}Ti_x)_3O_8)$ ($x$ = 0.4, 0.5, 0.6), $(Fe_2As_2)(Ba_4Sc_3O_{7.5})$, and $(Fe_2As_2)(Ba_3Sc_2O_5)$. Samples of $(Fe_2As_2)(Ba_3Sc_2O_5)$ showed semiconducting behaviors, while $(Fe_2As_2)(Sr_4(Sc_{1-x}Ti_x)_3O_8)$ and $(Fe_2As_2)(Ba_4Sc_3O_{7.5})$ showed metallic behaviors in their normal states. There were decreases in the resistivities of $(Fe_2As_2)(Sr_4(Sc_{0.6}Ti_{0.4})_3O_8)$ and $(Fe_2As_2)(Ba_4Sc_3O_{7.5})$ due to superconducting transition at ~30 and ~13 K, respectively. Chen *et al.* reported the possibility of superconductivity occurring in partially Ti-substituted $Sr_2ScFeAsO_3$ and $Sr_3Sc_2Fe_2As_2O_5$[9)]. Similarly, two-step resistivity transition was observed in $(Fe_2As_2)(Sr_4(Sc_{0.6}Ti_{0.4})_3O_8)$, although its origin has not been clarified. Although absolute values of resistivity cannot be compared quantitatively owing to quite low bulk density (~30% of the theoretical density), the temperature dependence of the normal-state resistivity of $(Fe_2As_2)(Sr_4(Sc_{1-x}Ti_x)_3O_8)$ samples becomes gradual by increasing the Ti composition. This indicates the difference in carrier concentrations of these samples.

New layered iron pnictide oxides $(Fe_2As_2)(Sr_4(Sc_{1-x}Ti_x)_3O_8)$, $(Fe_2As_2)(Ba_4Sc_3O_{7.5})$, and $(Fe_2As_2)(Ba_3Sc_2O_5)$ were synthesized and their physical properties examined. The structure of $(Fe_2As_2)(Sr_4(Sc_{1-x}Ti_x)_3O_8)$ and $(Fe_2As_2)(Ba_4Sc_3O_{7.5})$ consists of the alternate stacking of antifluorite $Fe_2As_2$ layers and triple perovskite-type oxide layers. These compounds are the first examples of this type of structure in layered iron pnictides. The bulk sample of $(Fe_2As_2)(Sr_4(Sc_{0.6},Ti_{0.4})_3O_8)$ showed diamagnetism up to 28 K and zero resistivity at 10 K. Different physical properties observed in $(Fe_2As_2)(Sr_4(Sc_{1-x}Ti_x)_3O_8)$ with $x$ = 0.4~0.6 indicated that the precise control of the cation composition might be crucial for the optimization of superconducting properties.

**Acknowledgments** This work was partly supported by a Grant-in-Aid for Young Scientists (B) No. 21750187, 2009, from the Ministry of Education, Culture, Sports, Science and Technology (MEXT), Japan, and the inter-university Cooperative Research Program of the Institute for Materials Research, Tohoku University.

**Figure captions**

Figure 1. Crystal structures of (a) 22325, (b) 22426, and (c) 22438.

Figure 2. Powder XRD pattern of $(Fe_2As_2)(Sr_4(Sc_{1-x}Ti_x)_3O_8)$ ($x$ = 0.4, 0.5, 0.6), $(Fe_2As_2)(Ba_4Sc_3O_{7.5})$, and $(Fe_2As_2)(Ba_3Sc_2O_5)$.

Figure 3. Temperature dependence of ZFC and FC magnetization curves of $(Fe_2As_2)(Sr_4(Sc_{0.6}Ti_{0.4})_3O_8)$, $(Fe_2As_2)(Sr_4(Sc_{0.5}Ti_{0.5})_3O_8)$, and $(Fe_2As_2)(Ba_4Sc_3O_{7.5})$ measured under 1 Oe and $(Fe_2As_2)(Sr_4(Sc_{0.4}Ti_{0.6})_3O_8)$ measured under 1000 Oe.

Figure 4. Temperature dependences of resistivity of (a) $(Fe_2As_2)(Sr_4(Sc_{1-x}Ti_x)_3O_8)$ ($x$ = 0.4, 0.5, 0.6), (b) $(Fe_2As_2)(Ba_4Sc_3O_{7.5})$, and (c) $(Fe_2As_2)(Ba_3Sc_2O_5)$.

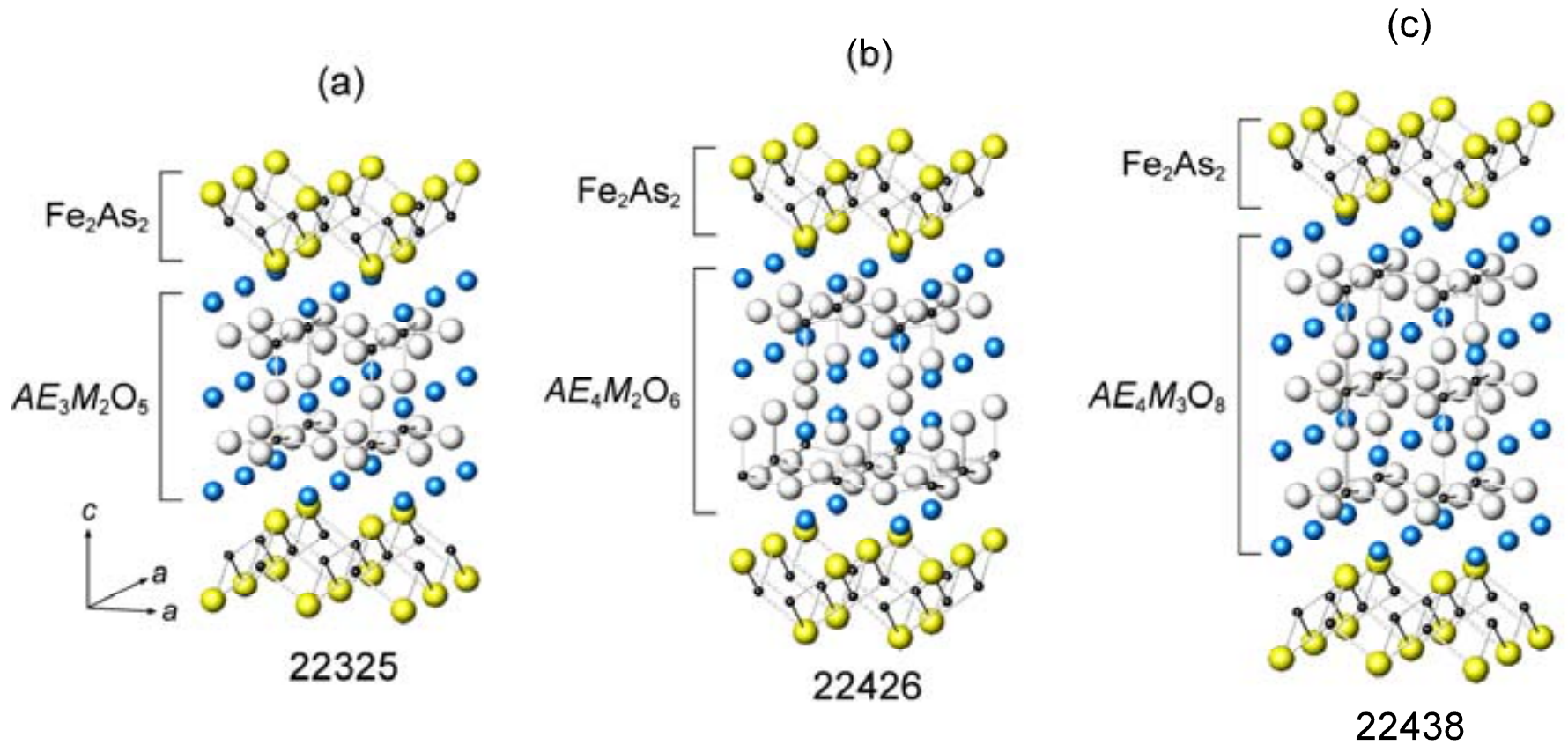

Figure 1

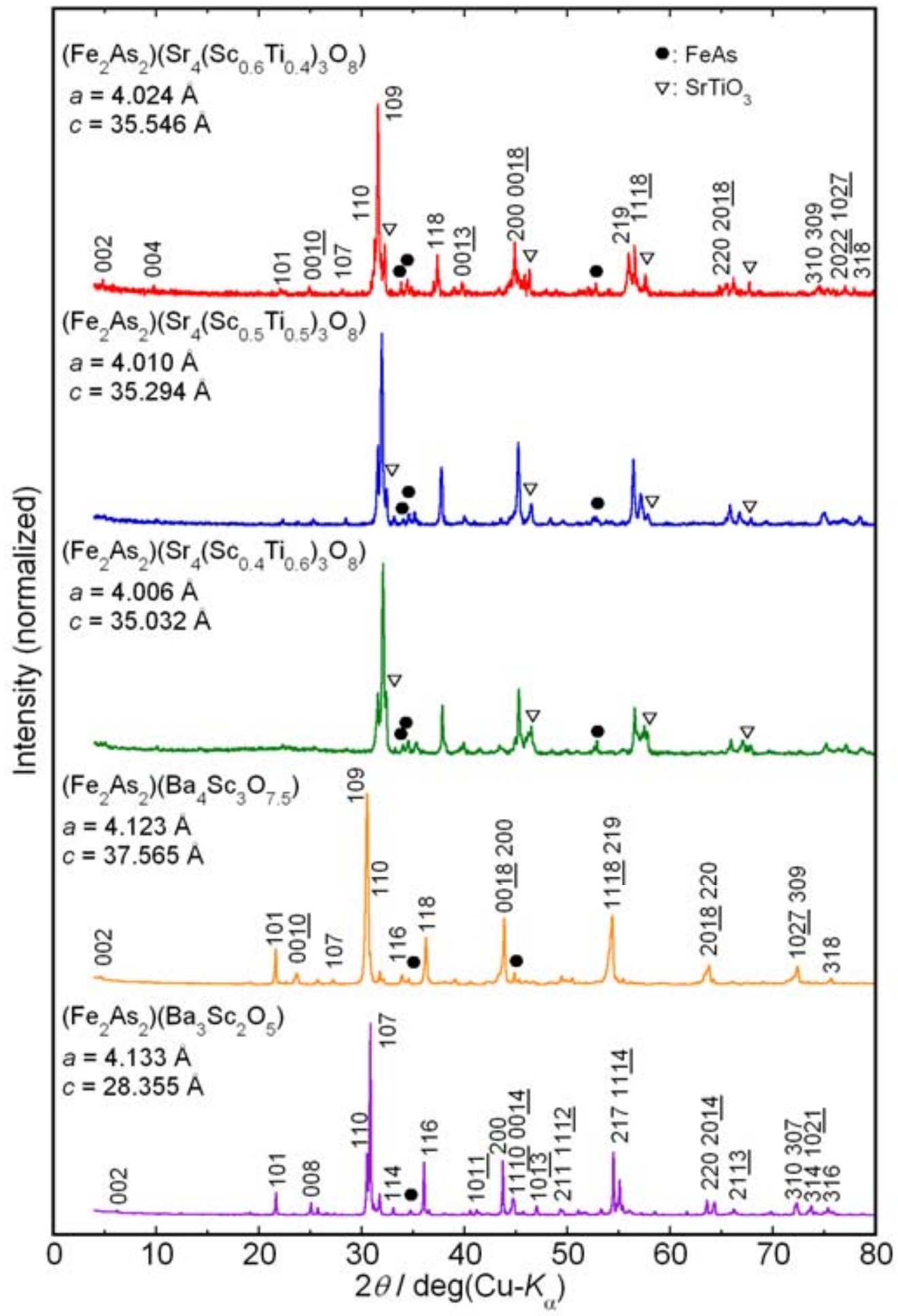

Figure 2

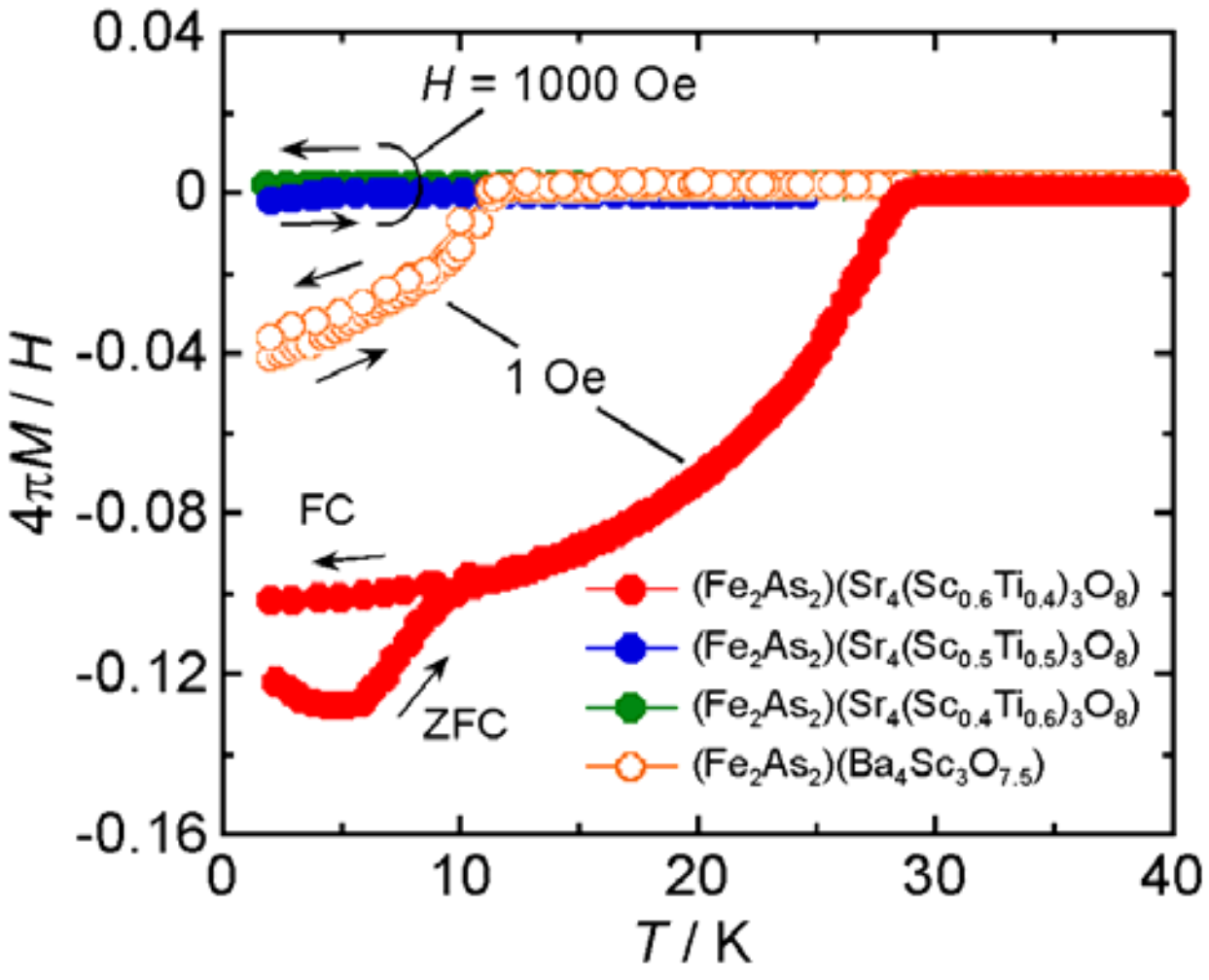

Figure 3

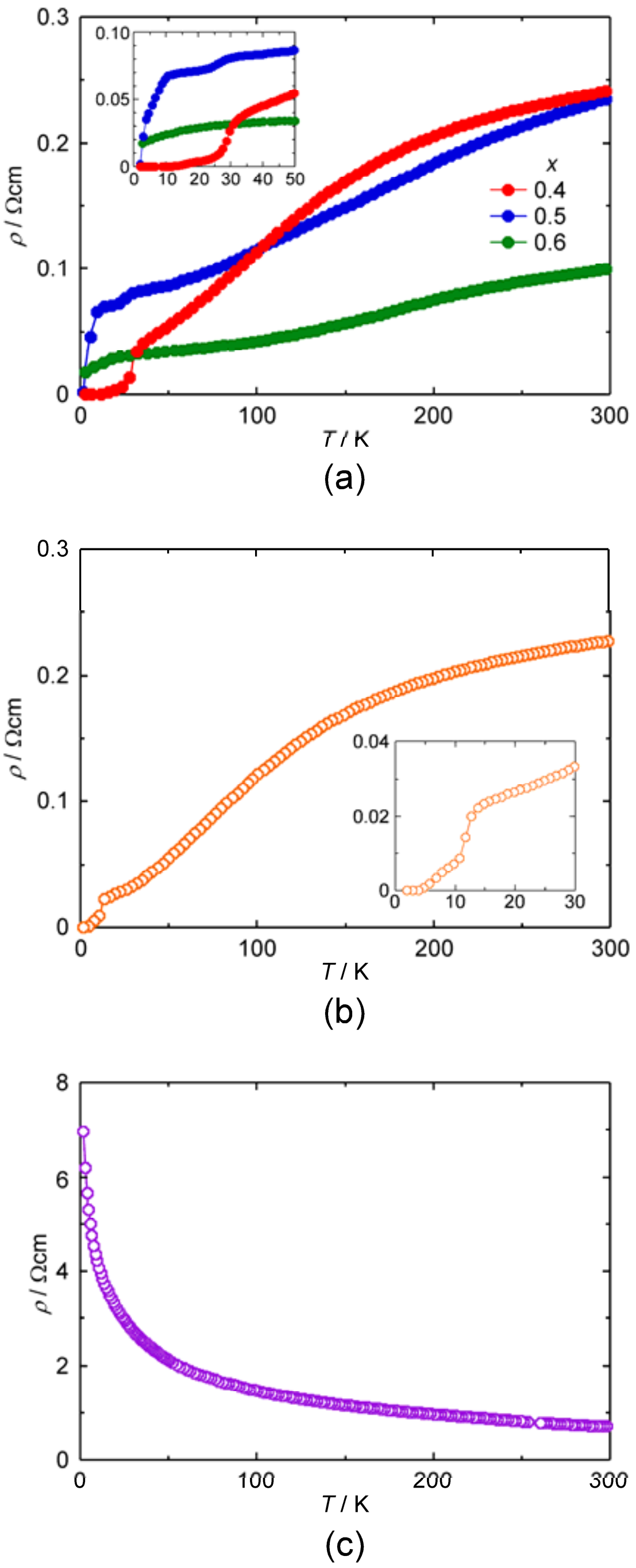

Figure 4